\documentclass[sigconf]{acmart}
\makeatletter
\@ACM@balancefalse
\makeatother

\AtBeginDocument{%
  }

\setcopyright{acmlicensed}
\copyrightyear{2018}
\acmYear{2018}
\acmDOI{XXXXXXX.XXXXXXX}
\acmConference[Conference acronym 'XX]{Make sure to enter the correct
  conference title from your rights confirmation email}{June 03--05,
  2018}{Woodstock, NY}
\acmISBN{978-1-4503-XXXX-X/2018/06}

\usepackage{listings}
\usepackage{algorithm}
\usepackage{algpseudocode}
\usepackage{tikz}
\usepackage{dsfont} 
\usepackage{placeins}
\usepackage{stfloats} 
\newcommand{\ind}{\mathord{\mathds{1}}}
\usetikzlibrary{positioning, arrows.meta, shapes.geometric, fit, calc, backgrounds, decorations.pathreplacing, patterns}

\AtBeginDocument{\raggedbottom}

\newtheorem{proposition}{Proposition}
\newtheorem{example}{Example}

\lstdefinestyle{idsql}{
  language=SQL,
  basicstyle=\small\ttfamily,
  keywordstyle=\bfseries,
  commentstyle=\color{gray},
  stringstyle=\color{black},
  showstringspaces=false,
  frame=single,
  framesep=4pt,
  xleftmargin=4pt,
  xrightmargin=4pt,
  aboveskip=6pt,
  belowskip=6pt,
  morekeywords={REGISTER,PREDICATE,ATTRIBUTE,SHAPLEY,OVER,WINDOW,ROWS,RANGE,PRECEDING,ORDER,DELTA,EMIT,CHANGES,APPROXIMATE,USING,RESERVOIR,SAMPLE,SHARE,LIFT}
}

\begin{document}
\setlength{\emergencystretch}{2em}
\hbadness=10000
\vbadness=10000
\raggedbottom

\title{Incremental Delta-Shapley: A Standalone Runtime for Predicate Attribution on Sliding Windows}

\author{Pouya Khani}
\affiliation{%
  \institution{Aarhus University}
  \city{Aarhus}
  \country{Denmark}
}
\email{pouya.khani@cs.au.dk}

\author{Ira Assent}
\affiliation{%
  \institution{Aarhus University}
  \city{Aarhus}
  \country{Denmark}
}
\email{ira@cs.au.dk}

\renewcommand{\shortauthors}{Khani et al.}

\begin{abstract}
Continuous aggregate queries over sliding windows are common in real-time analytics, but most systems report \emph{what} an aggregate is doing without attributing \emph{which} predicates account for the result. A companion paper~\cite{khani2026closedformpredicatelevelshapleyattribution} shows that exact predicate-level Shapley attribution for SUM, COUNT, AVG, and variance needs only three additive predicate summaries with closed-form coefficients. Those results settle the mathematics, not how a runtime maintains summaries across slides, exposes attribution, answers unregistered predicates, or amortizes repeated ad hoc ones.

We present \textbf{IDS} (Incremental Delta-Shapley), a standalone single-node runtime that turns those closed forms into a deployable explanation system. IDS consumes window-maintenance deltas, updates global, marginal, and atom summaries, and evaluates any closed form in constant time. Overlapping predicates use atomic refinement, and a restricted SQL-like API exposes attribution and its per-slide change as first-class operators. Unregistered predicates are answered by a retained-state scan, an inverted index, or an amortized sliding-window sample with concentration guarantees; frequent ones are promoted by rebuilding the refinement. On synthetic, adversarial, NEXMark-style, and NYC taxi workloads, attribution matches exhaustive Shapley enumeration to floating-point precision; incremental maintenance is flat in $N$ and up to $4.3\times10^{5}\times$ faster than per-window scans of the same form; and adaptive promotion cuts ad hoc cost by up to $9.2\times$ on Zipfian traces.
\end{abstract}



\maketitle

\section{Introduction}

The growth of high velocity data streams has driven the adoption of Data Stream Management Systems (DSMS). These streams appear in domains that range from algorithmic trading and IoT monitoring to real time cybersecurity. DSMS engines evaluate continuous aggregate queries over sliding windows at high rates. However, an aggregate alone does not tell an operator which part of the traffic changed it. Today, operators answer this explanation question out of band. They use handwritten drill down queries or export the window to a warehouse. In both cases, the answer arrives after the window that raised the question has expired.

The Shapley value~\cite{shapley1953value} provides an axiomatic distribution of the aggregate result among the input tuples. It is the unique value that satisfies Efficiency, Symmetry, Additivity, and the Dummy (null player) axiom. For generic cooperative games, however, exact Shapley computation is exponential. Even popular sampling based approximations such as KernelSHAP~\cite{lundberg2017unified} are stateless. They treat each window transition $W_t \to W_{t+1}$ as a fresh problem, although it typically shares $>99\%$ of its tuples with the previous window.\label{sec:intro-stateless} Recent closed form results~\cite{khani2026closedform} cover decomposable aggregates (SUM, COUNT) and sufficient statistic nonlinear aggregates (AVG, population and sample variance). They show that \emph{predicate level} Shapley attribution is an affine function of three additive predicate summaries. Here, predicate level attribution is the sum of tuple Shapley values over the matching coalition. The coefficients are closed forms in the running harmonic numbers $H_N$ and $H_N^{(2)}$. The formulas therefore settle \emph{what} to compute. They do not settle how a streaming runtime should maintain state, expose queries, answer unregistered predicates, or amortize promotion.

The remaining problem is a systems problem. To make predicate level Shapley attribution useful in practice, a runtime must decide:
\begin{itemize}
    \item how per predicate summaries are maintained across slides, how the active window's maintenance delta is obtained, and what state beyond the summaries must be retained;
    \item how attribution is surfaced to the analyst (a query interface, not a set of formulas);
    \item what to do when an analyst asks about a predicate that was not declared in advance (the \emph{ad hoc} case);
    \item how to amortize repeated ad hoc questions back into the registered set without corrupting the atomic refinement.
\end{itemize}

\begin{example}[Running Example: What the Operator Has to Type]
The companion paper motivates predicate attribution through an operations team
that monitors request latency for a cloud service~\cite{khani2026closedform}. We
reuse that scenario. Here, the question is not \emph{what} the attribution is,
because the closed forms answer that question. Instead, we ask what the operator
must do to obtain the attribution while the window is still active. Suppose
\textnormal{\texttt{region='EU-West'}} and \textnormal{\texttt{tier='premium'}} were declared to the engine
in advance. Each should support a constant time lookup. The same should hold for
\textnormal{\texttt{region='EU-West' AND NOT tier='premium'}}, although that combination was
never declared. Now suppose the operator suspects a specific service endpoint
that nobody registered. The engine must still answer from its available state.
It must also report which mechanism it used and the corresponding error. If the
operator asks about that endpoint a hundred times in the next minute, the engine
should stop paying the ad hoc cost. This paper addresses these three demands: a
declarative surface, a graceful ad hoc path, and amortization between them.
\end{example}

\subsection{Contributions}

We present \textbf{IDS} (Incremental Delta-Shapley), a standalone streaming attribution runtime. Once the closed forms of~\cite{khani2026closedform} exist, evaluating them on a static window is no longer the research question. The remaining question is how to turn those forms into a continuously maintained engine. IDS answers that question. Concretely:

\begin{itemize}
    \item \textbf{Attribution State Maintained by Deltas.} IDS is a single node runtime that consumes explicit maintenance deltas $(X^{out}_t,X^{in}_t)$ from a replay driver or window manager. It updates global summaries $(N,A(W),B(W))$, per predicate summaries, and atom summaries in mutable engine state. Maintenance across window slides is $\mathcal{O}(|\Delta_t|\,K)$ for $K$ registered predicates and $\mathcal{O}(|\Delta_t|)$ per predicate. IDS achieves this cost by observing insertions and expirations directly instead of recomputing each overlapping window instance from scratch.
    \item \textbf{Atomic Refinement with Marginal Summaries.} For $K\geq 2$ overlapping registered predicates, we maintain summaries at the atom level for compositional queries and, in parallel, marginal summaries for $\mathcal{O}(1)$ single predicate lookup. Arbitrary Boolean combinations of registered predicates are answered from matching active atoms; Efficiency is restored over the atom partition.
    \item \textbf{Declarative Query API.} We propose a restricted SQL-like surface with \texttt{REGISTER PREDICATE} together with \texttt{SHAPLEY\_ATTRIBUTE} and \texttt{SHAPLEY\_DELTA} as first class operators, with optional share--lift metadata for AVG so that variable-$N_t$ attributions remain interpretable~\cite{khani2026closedform}.
    \item \textbf{Three Ad Hoc Mechanisms.} For unregistered predicates, we present a retained state scan (exact, $\mathcal{O}(N)$), an optional inverted index (exact, depending on the posting list), and a sliding window uniform sample. The sample uses the chain-sample / priority-sample algorithms of Babcock et al.~\cite{babcock2002sampling}. We state the concentration bound for both regimes. Hoeffding applies to the sample with replacement that these algorithms produce natively. The tighter Hoeffding--Serfling form applies when the sample is a genuine size-$r$ draw without replacement.
    \item \textbf{Exact Predicate Promotion.} A frequency based policy migrates repeated ad hoc predicates into the registered set by scanning retained active window state and rebuilding the atomic refinement. A mode initialized from a sample is supported only as a clearly labeled provisional approximate bootstrap; its initial error does not decay under exact deltas alone.
    \item \textbf{Implementation and Evaluation.} We implement the design as a Python~3.12 + NumPy/Numba standalone library with a 165-test suite. The suite includes an exhaustive $N\le12$ Shapley oracle and executes the paper's listings verbatim. We evaluate the library by deterministic offline replay on four workloads. We measure summary cost separately from retained state, sampler, and index cost (Section~\ref{sec:eval}).
\end{itemize}

\paragraph{What the companion paper supplies, and what this paper adds.}
The companion paper~\cite{khani2026closedform} supplies the closed forms, their proofs, the atomic refinement \emph{theory}, the time based window generalization, and the boundary of the sufficient statistic phenomenon. Section~\ref{sec:background} restates only the two facts the engine consumes. Everything else that is reused is cited, not restated. This paper does \emph{not} claim those mathematical results. Its contribution begins after the formulas are known:
\begin{itemize}
    \item a \emph{delta state machine} that maintains global, marginal, and atom summaries under insertions and expirations;
    \item joint maintenance of atom and marginal summaries in the engine, including compaction and Efficiency over the live partition;
    \item a declarative SQL-like surface with registered, compositional, and delta operators;
    \item an ad hoc stack (scan, index, sample) with concentration bounds for sampling with replacement and without replacement, which the companion paper does not develop;
    \item exact promotion semantics that rebuild the refinement without corrupting registered state, plus a labeled provisional bootstrap whose error does not decay under deltas alone;
    \item a measured prototype that separates summary cost from retained window, sampler, and index cost.
\end{itemize}
In short, the companion paper answers \emph{what} $\Phi_{\mathcal{P}}$ equals; this paper answers \emph{how} a runtime maintains, queries, approximates, and amortizes that quantity online. The concrete operators cover the five aggregate games of~\cite{khani2026closedform}. The companion theory also characterizes every \emph{moment polynomial} game; IDS exposes that class as an extension interface with $\mathcal{O}(D)$ power sum state (Section~\ref{sec:extension}). Quantiles, MIN/MAX, joins, and model inference queries remain out of scope. Attribution here means cooperative game credit assignment over predicates, not causal root cause identification.

\begin{figure*}[tbp]
\centering
\includegraphics[width=\textwidth]{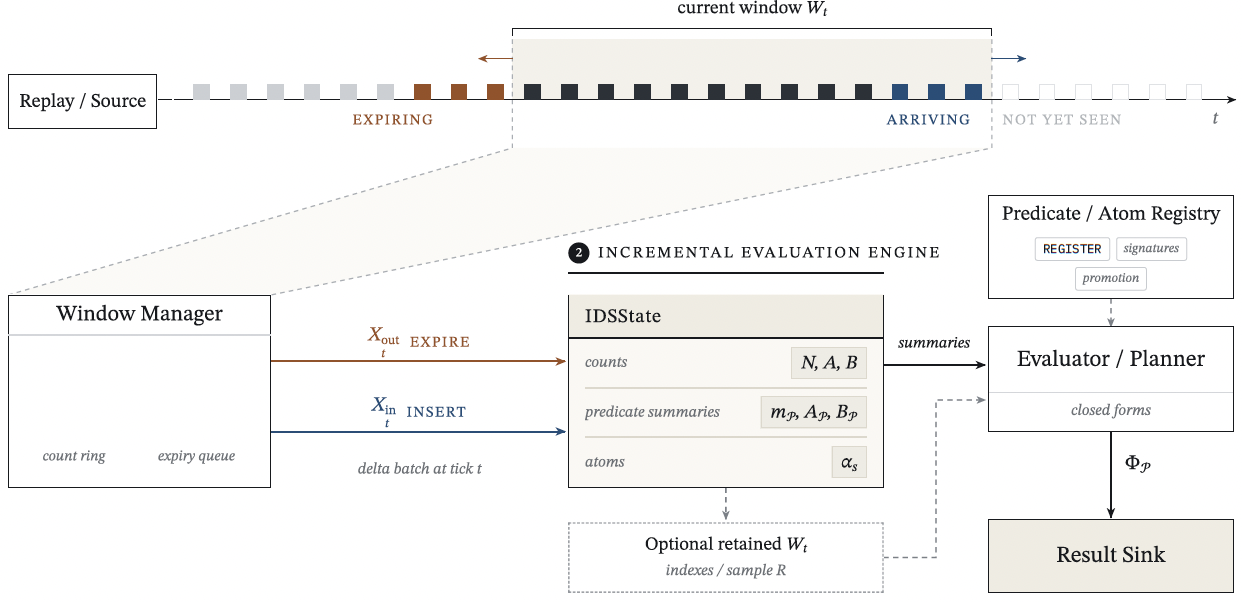}
\caption{Standalone IDS architecture. The Window Manager converts arrivals and expirations in $W_t$ into delta batches $(X^{out}_t,X^{in}_t)$, which update global, predicate, and atom summaries in \texttt{IDSState}. The registry and planner then select the registered, compositional, or ad hoc path, evaluate the closed forms~\cite{khani2026closedform}, and emit $\Phi_{\mathcal{P}}$ to the Result Sink.}
\Description{Architecture diagram showing a replay source feeding a sliding window manager. Expired and arriving tuples update global, predicate, and atom summaries in IDSState. A predicate registry and optional retained window state feed an evaluator, which sends predicate attribution to a result sink.}
\label{fig:architecture}
\end{figure*}

Figure~\ref{fig:architecture} gives an end to end view of the architecture.

\section{Background: What the Engine Evaluates}
\label{sec:background}

This section states only what is needed to read the rest of the paper. The
companion paper~\cite{khani2026closedform} establishes the closed forms, their
derivations, the atomic refinement theory, the time based window generalization,
and the boundary of the sufficient statistic phenomenon. We do not repeat their
proofs or supporting development here. Our notation follows that paper and is
summarized in Table~\ref{tab:notation}.

\subsection{Stream and Attribution Model}
Let $\mathcal{S}=\{x_1,x_2,\dots\}$ be an unbounded stream of tuples, each with a
timestamp $\tau_i$ and a numeric attribute $y_i$. A \emph{count based} window of
size $N$ holds the $N$ most recent tuples and slides by
$\Delta=(x_{out},x_{in})$; a \emph{time based} window of duration $\tau$ holds
the tuples with timestamp in $(t-\tau,t]$ and slides by a delta of multiple
tuples $\Delta_t=(X^{out}_t,X^{in}_t)$, with an active cardinality $N_t$ that varies
with the arrival process. Write $A(W)=\sum_{x_i\in W}y_i$ and
$B(W)=\sum_{x_i\in W}y_i^2$ for the window's first two raw power sums.

A predicate is a Boolean function $\mathcal{P}:\mathcal{D}\to\{0,1\}$ with active
coalition $C_{\mathcal{P}}=\{x\in W\mid\mathcal{P}(x)=1\}$ and summaries
$m_{\mathcal{P}}=|C_{\mathcal{P}}|$,
$A_{\mathcal{P}}=\sum_{C_{\mathcal{P}}}y_i$,
$B_{\mathcal{P}}=\sum_{C_{\mathcal{P}}}y_i^2$. Predicate attribution is the
sum of member values
$\Phi_{\mathcal{P}}(W)=\sum_{x_i\in C_{\mathcal{P}}}\phi_i(\nu,W)$, where
$\phi_i$ is the Shapley value of tuple $i$ for the aggregate game $\nu$ on the
grand coalition $W_t$~\cite{khani2026closedform}. We call $\mathcal{P}$
\emph{registered} if IDS maintains its summaries as the stream evolves and
\emph{ad hoc} otherwise. This distinction belongs to the engine, not the theory,
and it organizes the rest of the paper. The alternative that treats the predicate
as a player in a metagame is a different modeling choice, quantified
in~\cite{khani2026closedform}; IDS implements the sum of member values semantics
throughout.
\label{sec:overlap-alt}

\subsection{The Two Facts IDS Is Built On}
\label{sec:recap}

Everything in this paper rests on two results from~\cite{khani2026closedform}.

\paragraph{Fact 1 (closed forms).}
For SUM, COUNT, AVG, population variance, and sample variance,
$\Phi_{\mathcal{P}}(W)$ is a closed form in the global summaries
$(N,A(W),B(W))$, the predicate summaries
$(m_{\mathcal{P}},A_{\mathcal{P}},B_{\mathcal{P}})$, and the running harmonic
numbers $H_N$ and $H^{(2)}_N$~\cite{khani2026closedform}. No coalition
enumeration and no Shapley values at the tuple level are required. The companion
paper gives the explicit formulas, including AVG's share and lift decomposition
and the variance game coefficients. IDS implements those formulas and exposes
share and lift through its query API (Section~\ref{sec:sql}) when the analyst
requests them. A negative variance attribution denotes a stabilizing
contribution, not an error. We write $\Phi^{VAR}_{\mathcal{P}}$ and
$\Phi^{VAR\_SAMP}_{\mathcal{P}}$ for the two variance games.

\paragraph{Fact 2 (affinity), the structural fact this paper exploits.}
Once $N$, $A(W)$ and $B(W)$ are fixed, every one of the five closed forms is an
\emph{affine} function of the predicate summaries:
\begin{equation}
    \Phi_{\mathcal{P}}(W) \;=\; a_m\,m_{\mathcal{P}} + a_A\,A_{\mathcal{P}} + a_B\,B_{\mathcal{P}},
    \label{eq:affine}
\end{equation}
where $(a_m,a_A,a_B)$ depend only on the game, $N$, $A(W)$ and $B(W)$. The
companion paper derives the coefficients; IDS evaluates them from maintained
state.

Fact 2 is what makes IDS a system rather than a formula. It supports three
independent design decisions. First, additive state maintenance is exact
(Section~\ref{sec:state}). Second, summaries of disjoint groups may be added
before a single closed form evaluation. Compositional queries therefore need no
extra machinery (Section~\ref{sec:overlap}). Third, an unbiased estimate of the
three summaries yields an unbiased attribution estimate with concentration
guarantees. This result makes the sampling mechanism in
Section~\ref{sec:adhoc-reservoir} well founded rather than heuristic.

\begin{table}[tbp]
\normalsize
\renewcommand{\arraystretch}{1.12}
\centering
\begin{tabular}{|p{0.30\columnwidth}|p{0.60\columnwidth}|}
\hline
\textbf{Symbol} & \textbf{Meaning} \\ \hline
$W_t,N,N_t$ & Active window, fixed count window size, active time window size \\ \hline
$\Delta,\Delta_t$ & Count window slide of one tuple and time window slide of many tuples \\ \hline
$\mathcal{P},C_{\mathcal{P}},\Phi_{\mathcal{P}}$ & Predicate, its active coalition, and predicate attribution \\ \hline
$A(W),B(W)$ & Global sum and squared sum over $W$ \\ \hline
$m_{\mathcal{P}},A_{\mathcal{P}},B_{\mathcal{P}}$ & Predicate count, sum, and squared sum \\ \hline
$a_m,a_A,a_B$ & Affine attribution coefficients (Eq.~\ref{eq:affine}) \\ \hline
$H_N,H_N^{(2)}$ & First- and second order harmonic numbers \\ \hline
$\mathrm{sig}(x),\alpha_s,\mathcal{S}_W$ & Boolean signature, atom, and set of active signatures \\ \hline
$\mathcal{R}$ & Set of registered predicates \\ \hline
$R,r$ & Active window sliding sample and target sample size \\ \hline
$M$ & Bound on tuple values, $y_i\in[-M,M]$ (sampling analysis only) \\ \hline
\end{tabular}
\caption{Notation introduced or used by IDS. Symbols shared with the companion paper~\cite{khani2026closedform} keep their meaning there.}
\label{tab:notation}
\end{table}

\section{IDS Architecture}
\label{sec:architecture}

The IDS engine has four state components: attribution summaries maintained by deltas, a predicate/atom registry, optional retained active window state, and running harmonic numbers. The retained state supports exact ad hoc work, indexes, sampling, and promotion. Figure~\ref{fig:architecture} shows the standalone runtime pipeline: replay/source $\rightarrow$ window manager $\rightarrow$ delta batch $\rightarrow$ \texttt{IDSState} $\rightarrow$ evaluator/planner $\rightarrow$ result sink. We describe each component and then the slide algorithm.

\subsection{Attribution State Maintained by Deltas}
\label{sec:state}

IDS maintains the sufficient statistics needed by the closed forms in mutable engine state owned by the runtime. The required interface is the window maintenance delta: insertions $X^{in}_t$, expirations $X^{out}_t$, the current cardinality $N$ (or $N_t$), and the ability to update state when those events occur.

\textbf{How the delta is obtained.} In the standalone prototype, a \emph{window manager} derives $\Delta_t$ from a replay driver or live source. For count based windows, a ring buffer emits one expiration and one insertion per slide. For time based windows, a expiry queue ordered by timestamp emits a delta of multiple tuples at each emit instant. Alternatively, an upstream changelog may already provide signed inserts and deletes. The first two paths derive expiry locally and therefore retain $\mathcal{O}(N)$ active window state. Only the changelog path avoids this state, and only when no other feature requires the active tuples. Such features include exact ad hoc scans, indexes, and exact promotion. We report summary memory and total runtime memory separately (Section~\ref{sec:rq3}).

Each slide updates:
\begin{itemize}
    \item the global moments $N$ (or $N_t$), $A(W)$, $B(W)$;
    \item for each registered predicate $\mathcal{P}\in\mathcal{R}$, the marginal summaries $m_{\mathcal{P}}$, $A_{\mathcal{P}}$, $B_{\mathcal{P}}$ (for $\mathcal{O}(1)$ single predicate lookup);
    \item under $K\geq 2$ overlapping registered predicates, summaries at the atom level indexed by Boolean signature (Section~\ref{sec:overlap});
    \item the running harmonic numbers $H_N$ and $H_N^{(2)}$ when cardinality changes;
    \item optionally, a sliding window sample $R$, inverted indexes maintained by IDS, and the retained expiry structure.
\end{itemize}
The Shapley formula is evaluated only after the active window summaries are assembled; the system never merges local Shapley vectors.

\subsection{Predicate Registry}
The predicate registry maps each registered predicate $\mathcal{P}$ to the summary slots that store $m_{\mathcal{P}}$, $A_{\mathcal{P}}$, and $B_{\mathcal{P}}$. A predicate is registered explicitly via a \texttt{REGISTER PREDICATE} statement (Section~\ref{sec:sql}) or by the promotion policy (Section~\ref{sec:promotion}). The registry stores compiled membership functions $\mathcal{P}(x)$ that the slide loop calls in $\mathcal{O}(1)$ per tuple per predicate (or as a $K$-bit mask when atoms are maintained).

When the system maintains $K\geq 2$ registered predicates, the registry additionally tracks the active signature set $\mathcal{S}_W\subseteq\{0,1\}^K$ and the atom summary table indexed by signature.

\subsection{The Slide Loop}

\begin{algorithm}[t]
\caption{IDS slide update for registered predicates}
\label{alg:ids-update}
\begin{algorithmic}[1]
\State \textbf{Input:} delta $\Delta_t=(X^{out}_t,X^{in}_t)$ (for count based slides, take $X^{out}_t=\{x_{out}\}$ and $X^{in}_t=\{x_{in}\}$), registered predicates $\mathcal{R}$
\State Let $\sigma(x)=-1$ for $x\in X^{out}_t$ and $\sigma(x)=+1$ for $x\in X^{in}_t$ \Comment{signed multiset}
\ForAll{tuples $x$ in the signed multiset $X^{out}_t \uplus X^{in}_t$}
    \State $N \mathrel{+}= \sigma(x)$; \; $A(W) \mathrel{+}= \sigma(x)y$; \; $B(W) \mathrel{+}= \sigma(x)y^2$
    \State Form $\mathrm{sig}(x)$ over $\mathcal{R}$ in $\mathcal{O}(K)$ (bit mask) and apply the same signed update to atom $\alpha_{\mathrm{sig}(x)}$
    \ForAll{$\mathcal{P} \in \mathcal{R}$ with $\mathcal{P}(x)=1$}
        \State Apply the signed update $\sigma(x)\cdot(1,y,y^2)$ to the marginal $(m_{\mathcal{P}},A_{\mathcal{P}},B_{\mathcal{P}})$
    \EndFor
    \State (Optional): update sliding sample $R$, inverted indexes, and retained expiry state
\EndFor
\If{$N$ (or $N_t$) changed since last evaluation}
    \State Update running $H_N$, $H_N^{(2)}$ by $|\Delta N|$ steps and refresh the attribution coefficients \Comment{$\mathcal{O}(|\Delta N|)\le\mathcal{O}(|\Delta_t|)$}
\EndIf
\State On query, evaluate the requested closed form attribution from assembled summaries
\end{algorithmic}
\end{algorithm}

Each summary is a signed additive accumulator. A tuple entering the window adds $(1,y,y^2)$ to every summary whose predicate it satisfies. A tuple leaving the window subtracts the same triple~\cite{khani2026closedform}. The rule is identical for count based and time based deltas, so the cost is $\mathcal{O}(|\Delta_t|)$ per affected predicate in both cases. Algorithm~\ref{alg:ids-update} combines this rule with global summary updates, optional sample/index maintenance, and incremental harmonic updates. Harmonic numbers require no precomputation: $H_{N+1}=H_N+1/(N+1)$ and $H^{(2)}_{N+1}=H^{(2)}_N+1/(N+1)^2$. When $N_t$ jumps, IDS advances or retreats the pair by $|\Delta N_t|$ steps. The same tick's summary work already dominates this cost. Optional lazy prefix arrays up to $N^{\mathrm{seen}}$ are only an implementation convenience, not a correctness requirement.

\paragraph{Numerical stability.}
Long running streams can cause cancellation in $YT_{\mathcal{P}}$ and $Q_{\mathcal{P}}$ when the mean is much larger than the spread. Repeated additive updates can also cause drift~\cite{khani2026closedform}. IDS centers values as $y_i\leftarrow y_i-c$ for a coarse location estimate $c$. VAR attribution is shift invariant, and IDS recovers AVG attribution by adding back $c\,m_{\mathcal{P}}/N$. The hot path accumulators use compensated summation. IDS continuously checks the Efficiency identity $\sum_{s\in\mathcal{S}_W}\Phi_{\alpha_s}=\nu(W)-\nu(\emptyset)$ as a correctness probe.

\section{Maintaining the Atomic Refinement}
\label{sec:overlap}

With $K\geq2$ registered predicates, overlap makes the family
$\{\Phi_{\mathcal{P}_k}\}$ count some tuples more than once. Per predicate
summaries alone also cannot answer compositional queries such as intersections,
unions, complements, and set differences. The companion paper resolves both
problems through the \emph{atomic refinement}. Each tuple carries a $K$-bit
signature $\mathrm{sig}(x)$, and the atoms $\alpha_s$ partition $W$. Any Boolean
combination $\mathcal{Q}$ is the disjoint union of the atoms that satisfy it.
Therefore,
\begin{equation}
    \Phi_{\mathcal{Q}}(W) = \sum_{s\in\mathcal{S}(\mathcal{Q})\cap\mathcal{S}_W} \Phi_{\alpha_s}(W),
    \label{eq:compositional_query}
\end{equation}
where $\sum_{s\in\mathcal{S}_W}\Phi_{\alpha_s}=\nu(W)-\nu(\emptyset)$ restores
Efficiency over the induced partition. Per tick maintenance is
$\mathcal{O}(|\Delta_t|K)$, and state is $\mathcal{O}(|\mathcal{S}_W|)\le
\mathcal{O}(\min(2^K,N))$~\cite{khani2026closedform}. We take that theory as
given. This section records only what the \emph{engine} must decide in addition.

\paragraph{Marginals alongside atoms.}
Eq.~\ref{eq:compositional_query} answers a single registered predicate as a sum
over all atoms with bit $k$ set, which costs $\mathcal{O}(|\mathcal{S}_W|)$.
Yet single predicate queries are the common case in a monitoring workload. IDS
therefore maintains, in parallel with the atom table, a marginal summary triple
per registered predicate, updated in the same pass over $\Delta_t$. This makes
the common query $\mathcal{O}(1)$ instead of $\mathcal{O}(|\mathcal{S}_W|)$ at
the cost of $K$ extra triples, which Section~\ref{sec:rq3} shows is negligible.
Section~\ref{sec:rq2} measures both paths.

\paragraph{Merge before evaluating.}
Because attribution is affine in the summaries (Eq.~\ref{eq:affine}), a
compositional query has two exact evaluation strategies: evaluate the closed form
per matching atom and sum the results, or add the matching atom summaries first
and evaluate once. IDS implements the second by default, using one evaluation
instead of $|\mathcal{S}(\mathcal{Q})\cap\mathcal{S}_W|$, and retains the
first as a cross check; the test suite asserts they agree.

\paragraph{Compaction and the state bound.}
Signatures are allocated lazily and atoms whose count reaches zero are reclaimed,
so the resident table tracks $|\mathcal{S}_W|$ rather than $2^K$. When
$|\mathcal{S}_W|$ would still be too large, IDS falls back to materializing only
explicitly enumerated intersections; the remaining combinations then require a
retained state scan or an approximation, since inclusion--exclusion cannot invent
joint summaries that were never maintained. The actual growth of
$|\mathcal{S}_W|$ depends on the predicate set rather than on $K$. Both regimes
occur in practice. The companion paper reports that $|\mathcal{S}_W|$ reaches
its other bound $N$ for $K=64$ predicates drawn over cardinality-8 dimensions.
In contrast, the mutually exclusive categorical predicates of
Section~\ref{sec:rq2} keep it at $12$ for the same $K$.

\section{Declarative Query API}
\label{sec:sql}

Analysts interact with IDS through a restricted SQL-like declarative interface implemented by the prototype. This interface does not imply integration into an existing database or stream engine. It is the query surface that the standalone runtime compiles and executes. Analysts declare predicates with \texttt{REGISTER PREDICATE} and query them with two attribution operators. \texttt{SHAPLEY\_ATTRIBUTE} returns the current window attribution $\Phi_{\mathcal{P}}(W_t)$. \texttt{SHAPLEY\_DELTA} returns the change score $\Delta\Phi_{\mathcal{P}}(t)=\Phi_{\mathcal{P}}(W_{t+1})-\Phi_{\mathcal{P}}(W_t)$. Both operators are continuous and emit on every window slide, like the underlying aggregate. For AVG, an optional \texttt{WITH (SHARE, LIFT)} clause emits the share and lift components of the AVG closed form~\cite{khani2026closedform}. We recommend these components for interpreting attributions across time based windows with varying $N_t$.

Listing~\ref{lst:register} shows registration plus a continuous attribution query. Listing~\ref{lst:compositional} shows compositional predicate attribution over registered predicates. \texttt{SHAPLEY\_DELTA} is the consecutive-window change of $\Phi_{\mathcal{P}}$ and is used for alerting.

\begin{figure}[tbp]
\begin{lstlisting}[style=idsql, caption={Registering predicates and querying predicate level Shapley attribution for a sliding window AVG aggregate, with optional share--lift metadata.}, label={lst:register}]
-- Step 1: Register predicates of interest
REGISTER PREDICATE region_eu
    AS (region = 'EU-West')
    ON server_events;

REGISTER PREDICATE region_us
    AS (region = 'US-East')
    ON server_events;

-- Step 2: Continuous attribution query
SELECT
    SHAPLEY_ATTRIBUTE(
        AVG(latency), region_eu
        WITH (SHARE, LIFT)
    ) AS eu_attribution,
    SHAPLEY_ATTRIBUTE(
        AVG(latency), region_us
        WITH (SHARE, LIFT)
    ) AS us_attribution
FROM server_events
WINDOW w AS (
    ORDER BY event_time
    ROWS 10000 PRECEDING
)
EMIT CHANGES;
\end{lstlisting}
\Description{SQL-like example that registers EU-West and US-East predicates and continuously requests their AVG Shapley attributions together with share and lift metadata over a 10,000-row window.}
\end{figure}

\begin{figure}[tbp]
\begin{lstlisting}[style=idsql, caption={Compositional predicate attribution over registered predicates. Each combination is answered exactly from summaries at the atom level via Eq.~\ref{eq:compositional_query}.}, label={lst:compositional}]
SELECT
    SHAPLEY_ATTRIBUTE(
        AVG(latency),
        region_eu AND tier_premium
    ) AS eu_premium_attr,
    SHAPLEY_ATTRIBUTE(
        AVG(latency),
        region_eu AND NOT tier_premium
    ) AS eu_basic_attr,
    SHAPLEY_ATTRIBUTE(
        AVG(latency),
        region_eu OR region_us
    ) AS eu_or_us_attr
FROM server_events
WINDOW w AS (
    ORDER BY event_time
    ROWS 10000 PRECEDING
)
EMIT CHANGES;
\end{lstlisting}
\Description{SQL-like example that requests exact AVG attribution for an intersection, a set difference, and a union of registered predicates over a 10,000-row window.}
\end{figure}

\paragraph{Compilation.}
A \texttt{REGISTER PREDICATE} statement allocates a registry slot and compiles the predicate body into a membership function. For predicates registered before the stream starts, it initializes the summaries to zero. Otherwise, it bootstraps them through an exact retained state rebuild of the atomic refinement (Section~\ref{sec:promotion}). A \texttt{SHAPLEY\_ATTRIBUTE} call becomes a closed form evaluation against marginal or summaries at the atom level. For ad hoc predicates, the planner selects an ad hoc mechanism (Section~\ref{sec:adhoc}). \texttt{SHAPLEY\_DELTA} retains the previous attribution and subtracts it from the current attribution. It reports consecutive current-window changes, not a separate ``delta game''~\cite{khani2026closedform}.

\section{Ad Hoc Predicate Mechanisms}
\label{sec:adhoc}

Algorithm~\ref{alg:ids-update} maintains the summaries in $\mathcal{O}(|\Delta_t|\,K)$ per slide. Afterward, a registered single predicate query costs $\mathcal{O}(1)$, while a compositional query costs work proportional to the matching active atoms. In practice, debugging engineers also query predicates whose Boolean structure refers to attributes outside the registered set. We present three mechanisms for complementary regimes. The first is exact and always available. The second is exact and fast when an index applies. The third is approximate but uses the least state.

\subsection{Mechanism 1: Brute Force Retained State Scan}
\label{sec:adhoc-bf}
When an ad hoc predicate $\mathcal{P}$ arrives, scan the retained tuples in the active window, evaluate $\mathcal{P}(x)$ on each, and accumulate $(m_{\mathcal{P}},A_{\mathcal{P}},B_{\mathcal{P}})$. The global summaries $N$, $A(W)$, $B(W)$ are already maintained, so once the predicate summaries are known the closed form is $\mathcal{O}(1)$. Exact $\Phi_{\mathcal{P}}(W)$ is therefore computable in $\mathcal{O}(N)$ time given retained active window state. Retained state scans are exact but defeat the ``summaries only'' memory story for the registered path. They are the correctness fallback and the bootstrap path for exact promotion.

\subsection{Mechanism 2: Index Assisted Scan}
\label{sec:adhoc-idx}
IDS may optionally maintain inverted indexes on selected categorical attributes over the retained active window. These indexes are state owned by IDS with update and expiry cost $\mathcal{O}(|\Delta_t|\,|\mathcal{A}_{\mathrm{idx}}|)$ and memory $\mathcal{O}(N|\mathcal{A}_{\mathrm{idx}}|)$. For a conjunction of equality predicates on indexed attributes, exact $\Phi_{\mathcal{P}}(W)$ costs work proportional to the posting list scan plus summing the matches. For low selectivity indexed predicates this is faster than a full scan. For high selectivity predicates, the posting list work can be $\Theta(N)$. Disjunctions and ranges require appropriate index types; IDS dispatches accordingly.

\subsection{Mechanism 3: Sliding Window Sampling}
\label{sec:adhoc-reservoir}

IDS supports sampling based approximation for high throughput streams, large $N$, or settings without useful indexes. The mechanism uses the affine structure established in Section~\ref{sec:recap}. Once the global summaries are fixed, each closed form is an affine function of $(m_{\mathcal{P}},A_{\mathcal{P}},B_{\mathcal{P}})$. Estimating these three summaries from a sample is therefore enough to estimate the attribution. Because IDS maintains $N$, $A(W)$, and $B(W)$ exactly, the resulting plugin estimator inherits the unbiasedness of the summary estimators.

\paragraph{Sampler.}
IDS maintains a single active window sample $R$ of target size $r$ using the sliding window algorithms of Babcock, Datar, and Motwani~\cite{babcock2002sampling}: chain-sample for count based windows and priority-sample for time based windows. These algorithms maintain a genuine uniform sample over the currently active window. Running $r$ independent size-$1$ copies yields a sample of size $r$ \emph{with} replacement by default. A without replacement sample is available when the engine retains enough candidates (or uses a top-$r$ priority variant); only then does the finite population factor $\rho_r=1-(r-1)/N$ apply. Otherwise the concentration analysis uses $\rho_r=1$ (plain Hoeffding). We report sampler memory in addition to any state retained for expiry.

\paragraph{Estimators.}
On an ad hoc predicate query, IDS estimates
\begin{align}
    \widehat{m}_{\mathcal{P}}&=\frac{N}{|R|}\sum_{x_i\in R}\ind_{\mathcal{P}}(x_i),\label{eq:reservoir_m}\\
    \widehat{A}_{\mathcal{P}}&=\frac{N}{|R|}\sum_{x_i\in R}\ind_{\mathcal{P}}(x_i)\,y_i,\label{eq:reservoir_A}\\
    \widehat{B}_{\mathcal{P}}&=\frac{N}{|R|}\sum_{x_i\in R}\ind_{\mathcal{P}}(x_i)\,y_i^2,\label{eq:reservoir_B}
\end{align}
and substitutes them into the closed form. When each element of $R$ is a uniform draw from $W$, these estimators are unbiased for $(m_{\mathcal{P}},A_{\mathcal{P}},B_{\mathcal{P}})$ under sampling with or without replacement.

\paragraph{Stratified sampling.}
To protect rare categories, IDS may maintain independent sliding window samples within declared strata (region, service, tier) and apply the same plugin estimator within each stratum. The stratified estimators are unbiased whenever each nonempty stratum has a uniform sample. Stratification helps when the ad hoc predicate aligns with the strata; otherwise the index or retained state paths are preferable.

\subsection{Concentration Bounds}

\begin{proposition}[Sample Concentration for AVG Attribution]
\label{prop:reservoir}
Let $W$ be a window of size $N\geq2$ with values $y_i\in[-M,M]$, and let $R$ be a uniform sample of size $r\leq N$. Substituting Eqs.~\ref{eq:reservoir_m}--\ref{eq:reservoir_A} into the AVG closed form~\cite{khani2026closedform} yields an estimator $\widehat{\Phi}_{\mathcal{P}}^{AVG}$. Writing $\Phi_{\mathcal{P}}^{AVG}=\beta_1 A_{\mathcal{P}}+\beta_2 m_{\mathcal{P}}$ with coefficients fixed by the exactly maintained $N$ and $A(W)$, Hoeffding--Serfling~\cite{hoeffding1963probability,serfling1974probability,bardenet2015concentration} gives, with probability at least $1-\delta$,
\begin{equation}
    \left|\widehat{\Phi}_{\mathcal{P}}^{AVG}-\Phi_{\mathcal{P}}^{AVG}\right|
    \;\leq\;
    (|\beta_1|M+|\beta_2|)N
    \sqrt{\frac{2\rho_r\log(4/\delta)}{r}},
    \label{eq:reservoir_bound}
\end{equation}
where $\rho_r=1-(r-1)/N$ without replacement and $\rho_r=1$ with replacement. The leading factor is $\mathcal{O}(M\log N)$ and the rate is $\mathcal{O}(1/\sqrt{r})$. The variance games admit an analogous three-coordinate bound under Eq.~\ref{eq:affine}. Stratified samples replace the single deviation with a sum of per-stratum Hoeffding--Serfling terms.
\end{proposition}

\paragraph{Maintenance cost.}
A size-$r$ Babcock sample built from $r$ independent size-$1$ samplers costs $\mathcal{O}(|\Delta_t|\,r)$ per tick if domination counts are updated eagerly. That linear cost in $r$ is prohibitive at useful sample budgets, and a naive without replacement candidate rescan is worse.

IDS avoids this cost with \emph{amortized lazy pruning}. It appends each arriving tuple to the candidate list in $\mathcal{O}(1)$. When the list exceeds twice its expected steady state size $r+r\ln(N/r)$, a single newest-to-oldest sweep recomputes the survivor set with a size-$r$ min heap in $\mathcal{O}(|C|\log r)$ and removes at least half of the list, so the amortized cost is $\mathcal{O}(\log r)$ per arrival. The sampler still holds $\mathcal{O}(r\log(N/r))$ candidates in expectation. Stratified maintenance applies the same policy within each stratum.

\subsection{Mechanism Selection}
Table~\ref{tab:adhoc} summarizes the four regimes. With amortized pruning (Section~\ref{sec:adhoc-reservoir}), sampling costs under $1\,\mu$s per tuple and is flat in $r$ (Section~\ref{sec:rq4}). It is the only mechanism that need not retain the window. Section~\ref{sec:rq6} reports measured constants.

\begin{table*}[tbp]
\normalsize
\renewcommand{\arraystretch}{1.12}
\centering
\begin{tabular}{|l|c|c|c|c|}
\hline
\textbf{Mechanism} & \textbf{Query} & \textbf{Maint.} & \textbf{Memory} & \textbf{Error} \\ \hline
Brute force & $\mathcal{O}(N)$ & $\mathcal{O}(1)^\ast$ & $\mathcal{O}(N)$ retained & $0$ \\ \hline
Index assisted$^\dagger$ & $\mathcal{O}(I_{\mathcal{P}}+L_{\mathcal{P}}+|C_{\mathcal{P}}|)$ & $\mathcal{O}(|\Delta_t|\,|\mathcal{A}|)$ & $\mathcal{O}(N|\mathcal{A}|)$ & $0$ \\ \hline
Uniform sample & $\mathcal{O}(r)$ & $\mathcal{O}(|\Delta_t|\,r)^\ddagger$ & $\mathcal{O}(r)$ / $\mathcal{O}(r\log N)$ & Eq.~\ref{eq:reservoir_bound} \\ \hline
Stratified sample & $\mathcal{O}(\sum_h r_h)$ & $\mathcal{O}(|\Delta_t|\,r_{\max})$ & $\sum_h\mathcal{O}(r_h\log N_h)$ & Prop.~\ref{prop:reservoir} \\ \hline
\end{tabular}
\caption{Ad hoc predicate mechanisms. $|\mathcal{A}|$: attributes indexed by IDS; $r$: target sample size. $^\ast$ Beyond the retained window cost already paid for expiry. $^\dagger$ Conjunction of equality predicates on attributes indexed by IDS. $^\ddagger$ Reducible to $\mathcal{O}(|\Delta_t|(1+r/N))$ expected for chain-sample with hashed replacement indices and geometric skipping. Sampler memory is in addition to retained expiry state when present. Sampling error uses $\rho_r=1$ for with replacement samples and the Serfling factor without replacement.}
\label{tab:adhoc}
\end{table*}

\section{Adaptive Predicate Promotion}
\label{sec:promotion}

Ad hoc mechanisms are necessary for predicates that the operator did not anticipate, but each query carries a per call cost. When the same ad hoc predicate is asked repeatedly, IDS migrates it into the registered set.

\subsection{Promotion Policy}
IDS maintains a frequency table keyed by a canonical form of each ad hoc predicate. A predicate is promoted when its query count over a sliding observation window of length $T_{obs}$ exceeds a threshold $f_{prom}$. Conversely, a registered predicate that received no queries over a window of length $T_{stale}$ is demoted as stale.

\subsection{Bootstrapping the Summaries and Atoms}
Adding a predicate $P_{K+1}$ changes every existing $K$-bit signature. Exact promotion therefore cannot initialize only a marginal $(m,A,B)$ triple. It must split every active atom according to $P_{K+1}(x)$.

\begin{itemize}
    \item \textbf{Exact retained state bootstrap (default).} Scan the retained active window once, evaluate the new predicate on each tuple, repartition atom summaries, and initialize the new marginal. Cost $\mathcal{O}(N)$ paid once. Subsequent slides maintain the refined atoms exactly via Algorithm~\ref{alg:ids-update}.
    \item \textbf{Provisional sample bootstrap (approximate mode).} Initialize the new marginal from Eqs.~\ref{eq:reservoir_m}--\ref{eq:reservoir_B} and temporarily answer only marginal queries for $P_{K+1}$. Joint atoms involving $P_{K+1}$ remain unavailable until an exact rebuild. The bootstrap error $e=\widehat{T}_0-T_0$ is \emph{invariant} under subsequent exact inserts and deletes: $\widehat{T}_t-T_t=e$. IDS therefore treats this mode as provisional and schedules a background exact rebuild. It does not claim that the error decays under deltas alone.
\end{itemize}
Demotion of $P_k$ merges every pair of atoms that differed only in bit $k$, in $\mathcal{O}(|\mathcal{S}_W|)$ time.

\section{Standalone Prototype}
\label{sec:prototype}

We implemented IDS as a Python~3.12 + NumPy/Numba single node library. \texttt{IDSState} owns global and marginal summaries, the atom table, running harmonics, and optional retained expiry, index, and sample state. Window managers emit deterministic $\Delta_t$ batches; the predicate registry compiles membership functions and tracks promotion candidates; the planner dispatches registered, compositional, and ad hoc paths; a replay driver drives checksummed workloads; and a restricted DSL parser executes Listings~\ref{lst:register} and~\ref{lst:compositional} verbatim in the test suite. For RQ2 throughput, Numba-compiled slide loops isolate the algorithmic difference between incremental maintenance and a full window scan of the same summary set; they agree with \texttt{IDSState} to $10^{-9}$ relative. IDS also exposes a moment polynomial extension hook with $\mathcal{O}(D)$ power sum state~\cite{khani2026closedform}; the shipped operators remain SUM, COUNT, AVG, VAR\_POP, and VAR\_SAMP.
\label{sec:extension}

\section{Evaluation}
\label{sec:eval}

\paragraph{Setup.}
All experiments run single threaded on one Apple M2 Pro (10 cores, 16\,GiB RAM, macOS~15.5) under CPython~3.12.9 with NumPy~2.4 and Numba~0.66. The deterministic replay driver runs every configuration, so all configurations receive a byte identical delta sequence. Timing figures are means over five repetitions, and we report 95\% $t$-intervals in the result CSVs. We reran timing sensitive experiments with no other job on the machine. Table~\ref{tab:workloads} lists the workloads. Each workload is preprocessed once and fixed by a SHA-256 manifest.

\paragraph{Artifact.}
The library, the 165-test suite, the workload preparation scripts, one script per research question, and the figure/table generators are packaged with the paper. The generated CSV files are the source of record for the values written directly in the tables below. Reported cost units are chosen to be machine independent where possible: RQ5 counts \emph{tuples touched} rather than seconds.

\begin{table}[tbp]
\normalsize
\renewcommand{\arraystretch}{1.12}
\centering
\begin{tabular}{|l|r|l|l|}
\hline
\textbf{Workload} & \textbf{Tuples} & \textbf{Value $y$} & \textbf{Predicate attributes} \\ \hline
synthetic & 2\,000\,000 & latency & region, tier, endpoint \\ \hline
stress & 400\,000 & latency & region, tier, endpoint \\ \hline
nyc (real) & 2\,850\,723 & fare & borough, vendor, payment \\ \hline
auction & 1\,500\,000 & bid price & category, channel, auction \\ \hline
\end{tabular}
\caption{Workloads. \emph{synthetic} is a Poisson arrival lognormal latency stream carrying a planted incident (a $4\times$ latency shift on \texttt{region='EU-West' AND tier='premium'} over the middle 4\% of the stream). \emph{stress} is the adversarial numerical case $\mu=10^{9}$, $\sigma=10^{-3}$. \emph{nyc} is NYC TLC yellow taxi trips for January 2024, filtered to plausible fares and known boroughs. \emph{auction} is a NEXMark style bid stream with Zipfian auction popularity.}
\label{tab:workloads}
\end{table}

\subsection{RQ1: Correctness}
\label{sec:rq1}

The closed forms are exact by construction~\cite{khani2026closedform}. RQ1 asks whether the \emph{engine} preserves that exactness over long runs under delta maintenance, atom repartitioning, compensated accumulators, and checkpoint/restore. An enumeration oracle ($N\le12$) is a regression check. The main evidence is a \emph{delta equivalence} test over $3\times10^5$-slide replays on all four workloads, plus a continuous Efficiency probe $\sum_{s\in\mathcal{S}_W}\Phi_{\alpha_s}=\nu(W)-\nu(\emptyset)$ over the live atom partition.

\begin{table*}[tbp]
\normalsize
\renewcommand{\arraystretch}{1.12}
\centering
\begin{tabular}{|l|r|c|c|c|c|}
\hline
\textbf{Game} & \textbf{Cases} & \textbf{max abs err} & \textbf{max rel err} & \textbf{max rel delta err} & \textbf{max rel Eff.\ resid.} \\ \hline
SUM & 3\,186 & $3.6\!\times\!10^{-12}$ & $2.9\!\times\!10^{-12}$ & $3.1\!\times\!10^{-16}$ & $2.2\!\times\!10^{-16}$ \\
COUNT & 3\,186 & $1.7\!\times\!10^{-13}$ & $1.9\!\times\!10^{-14}$ & 0 & 0 \\
AVG & 3\,186 & $4.3\!\times\!10^{-14}$ & $6.1\!\times\!10^{-13}$ & $4.4\!\times\!10^{-15}$ & $5.4\!\times\!10^{-16}$ \\
VAR\_POP & 3\,186 & $1.4\!\times\!10^{-12}$ & $7.1\!\times\!10^{-13}$ & $1.4\!\times\!10^{-13}$ & $2.6\!\times\!10^{-15}$ \\
VAR\_SAMP & 3\,186 & $1.6\!\times\!10^{-12}$ & $2.7\!\times\!10^{-13}$ & $2.0\!\times\!10^{-13}$ & $3.0\!\times\!10^{-15}$ \\
\end{tabular}
\caption{RQ1, implementation validation. Columns 3--4 rerun the enumeration oracle of~\cite{khani2026closedform} as a regression check. Columns 5--6 are the tests specific to the engine: maintained by deltas versus recomputed from the window over $3\times10^5$-slide replays, and the Efficiency residual over the live atom partition. All errors are at the double precision floor.}
\label{tab:rq1}
\end{table*}

Table~\ref{tab:rq1} reports the result: every discrepancy is at the double precision floor ($\le 4\times10^{-12}$ absolute against enumeration, $\le 5\times10^{-13}$ relative for delta equivalence, $\le 4\times10^{-15}$ relative for the Efficiency residual). The $N\in\{0,1,2\}$ edge cases of~\cite{khani2026closedform} are covered by dedicated tests. The full suite is 165 tests and passes end to end; it also executes Listings~\ref{lst:register} and~\ref{lst:compositional} verbatim, so the query surface of Section~\ref{sec:sql} is an executable interface rather than an illustration.

This validation revealed two implementation constraints that are invisible in the mathematics. (i)~Under a count based window, priority samplers still require the expiry cutoff. Omitting it raises no error but silently admits expired tuples into the sample. (ii)~Chain-sample assumes a \emph{fixed} active count, so using it within a stratum is biased. Per stratum cardinalities $N_h$ vary even under a fixed size global window. Stratum samplers must therefore be variable window priority samplers.

\subsection{RQ2: Incremental throughput and attribution latency}
\label{sec:rq2}

The companion paper already establishes that closed form evaluation with incrementally maintained summaries is flat in $N$, while per window recomputation is not~\cite{khani2026closedform}. RQ2 asks how the advantage changes with slide size $|\Delta_t|$ and with $K$ registered predicates plus a full atom table. Both kernels are compiled and maintain the same summary set.

\begin{figure*}[tbp]
\centering
\includegraphics[width=0.95\linewidth,height=0.30\textheight,keepaspectratio]{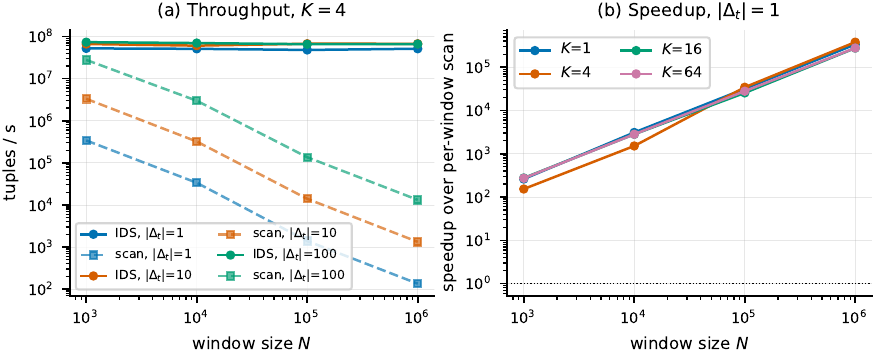}
\caption{RQ2. (a) Steady state ingest throughput versus window size for IDS and the per window scan baseline. IDS is flat in $N$; the baseline falls as $1/N$. (b) Speedup versus $N$ at $|\Delta_t|=1$ for each $K$.}
\Description{Two plots. The first shows IDS throughput remaining nearly constant as window size grows while scan throughput declines. The second shows IDS speedup increasing with window size for each tested predicate count.}
\label{fig:rq2}
\end{figure*}

Figure~\ref{fig:rq2} shows the results. IDS ingest throughput is \emph{flat in $N$}. Its median is $7.3\times10^{7}$ tuples/s at $|\Delta_t|=1$ and rises to $1.3\times10^{8}$ tuples/s at $|\Delta_t|=100$. The per window scan baseline degrades as $\Theta(1/N)$. Across all measured cells, speedup ranges from $2.6\times$ ($N=10^3$, $|\Delta_t|=100$, AVG) to $4.3\times10^{5}\times$ at $N=10^6$, $|\Delta_t|=1$. At the midrange point $N=10^5$, $K=4$, $|\Delta_t|=10$, the speedup is $4.8\times10^{3}$. Incremental maintenance wins whenever $N$ exceeds the slide size by more than a small constant.

Table~\ref{tab:rq2_latency} reports analyst-facing query latency through the Python API. Single predicate attribution from marginal summaries costs $1.38\,\mu$s at p50 and stays flat from $K=1$ to $K=64$. Compositional queries cost $1.6$--$4.0\,\mu$s at p50 and grow with $|\mathcal{S}_W|$, not with $K$. Active signatures reach $12$ at both $K=16$ and $K=64$ because the predicates are mutually exclusive categoricals, so the nominal $2^{64}$ atom bound is far from tight.

\begin{table}[tbp]
\normalsize
\setlength{\tabcolsep}{3pt}
\renewcommand{\arraystretch}{1.12}
\centering
\begin{tabular}{|r|r|l|r|r|}
\hline
$K$ & $|\mathcal{S}_W|$ & \textbf{path} & \textbf{p50} & \textbf{p99} \\ \hline
1 & 2 & marginal & 1.37 & 3.83 \\
1 & 2 & compositional & 1.58 & 3.83 \\
4 & 6 & marginal & 1.37 & 1.87 \\
4 & 6 & compositional & 2.54 & 4.04 \\
16 & 12 & marginal & 1.37 & 4.12 \\
16 & 12 & compositional & 3.92 & 10.21 \\
64 & 12 & marginal & 1.42 & 3.88 \\
64 & 12 & compositional & 4.00 & 8.21 \\
\end{tabular}
\caption{RQ2, library attribution latency ($\mu$s) at $N=10^4$. The marginal path is flat in $K$; the compositional path scales with active signatures.}
\label{tab:rq2_latency}
\end{table}

\subsection{RQ3: Memory accounting}
\label{sec:rq3}

\begin{figure*}[tbp]
\centering
\includegraphics[width=0.95\linewidth,height=0.30\textheight,keepaspectratio]{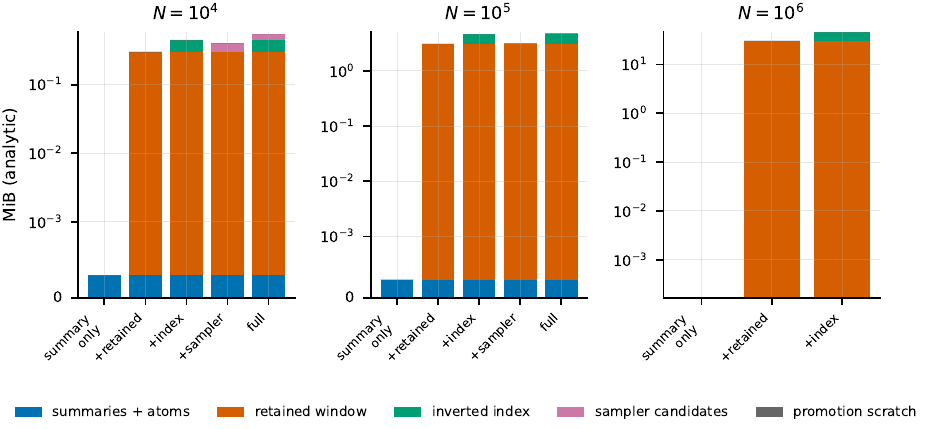}
\caption{RQ3. Analytic memory by component across configurations and window sizes (symmetric log scale). Summary and atom state is invisible at this scale by $N=10^5$.}
\Description{Memory breakdown plots across IDS configurations and window sizes. Retained window and index state grow with window size, while summary and atom state remains nearly invisible at larger window sizes.}
\label{fig:rq3}
\end{figure*}

Figure~\ref{fig:rq3} separates the memory components. Summary and atom state is $312$\,B at $K=4$ and is \emph{constant in $N$} across three orders of magnitude. Every other component is linear in $N$. At $N=10^5$, summaries are $0.006\%$ of the full analytic footprint; at $N=10^6$ with an index, they are $6\times10^{-4}\%$, while retained window state dominates at $30.5$\,MiB and the index adds $15.3$\,MiB. A summaries only configuration that receives deltas from an upstream changelog allocates no measurable heap at $N\le10^5$. The CPython peak for the full $N=10^5$ configuration is $44$\,MiB versus $4.6$\,MiB analytic, which is object overhead a compiled engine would not pay. Attribution state itself is essentially free; $\mathcal{O}(N)$ cost is paid only for exact ad hoc answers, indexes, or locally derived expiry.

\subsection{RQ4: Approximation quality and bound validity}
\label{sec:rq4}

\begin{figure*}[tbp]
\centering
\includegraphics[width=0.95\linewidth]{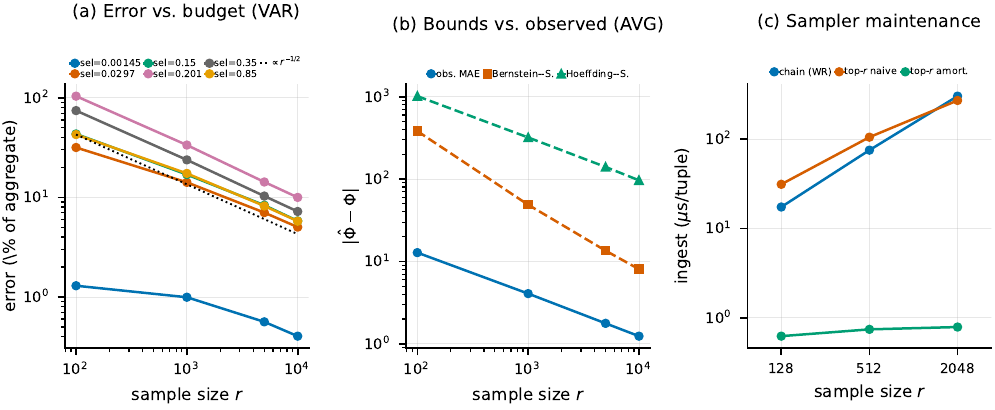}
\caption{RQ4. (a) Error as a percentage of the aggregate versus sample size, matching the $r^{-1/2}$ reference. (b) The two concentration bounds against the observed error; the empirical Bernstein--Serfling form is roughly an order of magnitude tighter. (c) Sampler ingest cost: amortized pruning is flat in $r$ where the eager and chain based variants grow linearly.}
\Description{Three sampling plots showing attribution error decreasing with sample size, empirical Bernstein--Serfling bounds tighter than Hoeffding--Serfling bounds, and amortized sampler maintenance remaining nearly flat as the sample size grows.}
\label{fig:rq4}
\end{figure*}

Section~\ref{sec:adhoc-reservoir} promises an $\mathcal{O}(1/\sqrt{r})$ error rate, a usable concentration bound, and cheap ingest. Figure~\ref{fig:rq4} tests all three. We normalize error by the aggregate being explained~\cite{khani2026closedform}, not by $\Phi_{\mathcal{P}}$. A control variate that regresses sample attribution contributions on the exactly maintained $(y,y^2)$ totals reduces variance by a median $1.06\times$ (up to $4.96\times$).

Figure~\ref{fig:rq4}(a) confirms the $\mathcal{O}(1/\sqrt{r})$ rate with bias indistinguishable from zero. Sampling $1\%$ of a $10^5$-tuple window estimates AVG attribution to within $0.4$--$7\%$ of the aggregate; $5\%$ sampling brings that to $0.2$--$3.1\%$. The variance game is harder ($1$--$14\%$ at $r/N=0.05$) because it depends on a second moment.

Figure~\ref{fig:rq4}(b) shows that the range based Hoeffding--Serfling bound is valid but loose. Replacing it with the empirical Bernstein--Serfling inequality of Bardenet and Maillard~\cite{bardenet2015concentration} tightens the bound by a median $22.5\times$ and reduces slack against the observed 95th-percentile error from $246\times$ to $10\times$, with empirical coverage $1.00$ against a nominal $0.95$. IDS reports the empirical Bernstein form by default.

Figure~\ref{fig:rq4}(c) shows why amortized lazy pruning decides whether sampling is usable. Eager domination updates cost $31$--$271\,\mu$s per tuple and grow with $r$. Amortized pruning costs $0.62$--$0.79\,\mu$s per tuple and is essentially flat in $r$ ($343\times$ cheaper at $r=2048$), without changing what is sampled. At realistic fractions $r/N\le0.1$, with replacement and without replacement MAE differ by at most $7\%$, so the scheme distinction is mainly a correctness requirement for which bound factor $\rho_r$ applies. For a stratum aligned rare predicate, stratified sampling cuts MAE by $1.3$--$1.4\times$ and reduces all-zero samples at $r=100$ from $10.5\%$ to $1.25\%$ of trials.

\subsection{RQ5: Adaptive promotion}
\label{sec:rq5}

We run a Zipfian ad hoc query trace against a $N=5\times10^4$ window ($1600$ queries from $40$ NYC predicates outside the registered set) and count \emph{tuples touched}. An ad hoc scan touches $N$ tuples; a registered lookup touches none; one exact promotion rebuild touches $N$ once.

Table~\ref{tab:rq5} shows that adaptive promotion falls between static registration and the oracle. At $\alpha=2.0$, static costs $80.0\times10^6$ tuples touched; adaptive with $f_{prom}=10$ costs $8.7\times10^6$ ($9.2\times$ reduction), while the oracle floor is $4.3\times10^6$. Rebuild overhead is under $5\%$ of the adaptive total. Exact promotion is $\mathcal{O}(N)$ ($\approx1\,\mu$s per tuple); demotion merges atom pairs in tens of microseconds independent of $N$. A provisional sample bootstrap's initial error is bit identical under exact deltas and drops to zero only after an exact rebuild.

\begin{table}[tbp]
\small
\setlength{\tabcolsep}{2.5pt}
\renewcommand{\arraystretch}{1.08}
\centering
\begin{tabular}{|r|l|r|r|r|r|}
\hline
$\alpha$ & \textbf{policy} & \textbf{prom.} & \textbf{ad hoc} & \textbf{rebuild} & \textbf{vs.\ static} \\
 & & & \textbf{($10^6$)} & \textbf{($10^6$)} & \\ \hline
1 & static & 0 & 80.00 & 0.00 & 1.00$\times$ \\
1 & oracle & 0 & 29.85 & 0.00 & 2.68$\times$ \\
1 & adaptive(f=10) & 8 & 35.15 & 0.40 & 2.25$\times$ \\
2 & static & 0 & 80.00 & 0.00 & 1.00$\times$ \\
2 & oracle & 0 & 4.25 & 0.00 & 18.82$\times$ \\
2 & adaptive(f=10) & 8 & 8.25 & 0.40 & 9.25$\times$ \\
\end{tabular}
\caption{RQ5. Zipfian query trace, $N=5\times10^4$, at most eight promotions. Adaptive lands between static and oracle.}
\label{tab:rq5}
\end{table}

\subsection{RQ6: Stateless Shapley and attribution for debugging}
\label{sec:rq6}

Permutation Monte Carlo~\cite{castro2009polynomial} estimates the same quantity as IDS. Grouped KernelSHAP~\cite{lundberg2017unified} on the two-player predicate game estimates a different \emph{metagame} value~\cite{khani2026closedform}. We score each baseline against its own estimand.

\begin{table}[tbp]
\centering
\scriptsize
\setlength{\tabcolsep}{2.5pt}
\renewcommand{\arraystretch}{1.08}
\resizebox{\columnwidth}{!}{%
\begin{tabular}{|r|l|r|l|r|}
\hline
$N$ & \textbf{method} & \textbf{lat.\ ($\mu$s)} & \textbf{rel.\ err} & \textbf{vs.\ IDS} \\ \hline
$10^{3}$ & IDS (closed form) & 1.3 & -- & 1$\times$ \\
$10^{3}$ & perm.\ MC (200) & 85\,190 & 8.43e-01 & 66\,264$\times$ \\
$10^{3}$ & KernelSHAP (100) & 296 & 2.67e-01 & 230$\times$ \\
$10^{4}$ & IDS (closed form) & 1.3 & -- & 1$\times$ \\
$10^{4}$ & perm.\ MC (200) & 1\,752\,013 & 1.06e-01 & 1\,303\,601$\times$ \\
$10^{4}$ & KernelSHAP (100) & 1\,391 & 2.68e-01 & 1\,035$\times$ \\
\end{tabular}%
}
\caption{RQ6, per window attribution cost. IDS uses summaries only; baselines hold the full window. Relative error is against each method's own estimand.}
\label{tab:rq6}
\end{table}

Table~\ref{tab:rq6} shows the gap. IDS answers in $1.3$--$1.4\,\mu$s, independent of $N$. Permutation Monte Carlo with 200 permutations needs $1.8\times10^{6}\,\mu$s at $N=2\times10^4$ to reach $11\%$ relative error and must hold the whole window. The gap widens with $N$, which is concrete evidence for the ``stateless'' critique in Section~\ref{sec:intro-stateless}.

\begin{figure}[tbp]
\centering
\includegraphics[width=\linewidth]{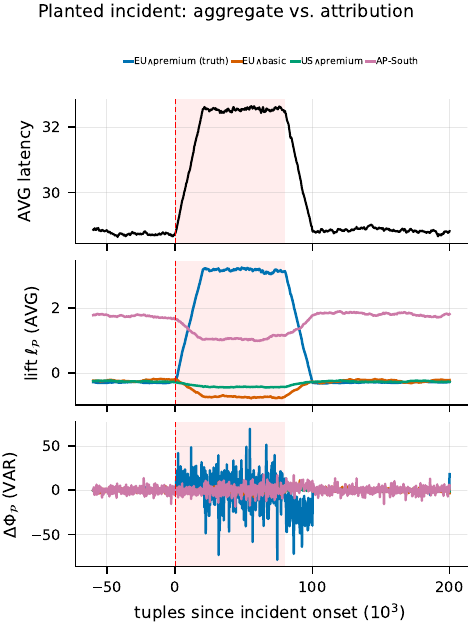}
\caption{RQ6 case study (attribution/alerting). Top: AVG around the planted change. Middle: AVG lift. Bottom: $\Delta\Phi^{VAR}$. The planted predicate ranks first on $\Delta\Phi$ within $79$ tuples of onset.}
\Description{Three aligned time series panels around a shaded planted change showing AVG, lift, and variance attribution ranking.}
\label{fig:rq6case}
\end{figure}

\paragraph{Case study: planted incident (attribution and alerting, not causation).}
This case study asks whether maintained attribution can \emph{rank} a known planted predicate among registered candidates soon after onset. It does \emph{not} claim causal root cause diagnosis. In the synthetic workload, a $4\times$ latency shift is planted on \texttt{region='EU-West' AND tier='premium'}; the ground truth column is removed before the engine sees any tuple. Four candidates are registered: the true predicate and three decoys. Figure~\ref{fig:rq6case} shows the result at $N=2\times10^4$. Within $79$ tuples of onset, $\Delta\Phi^{VAR}$ ranks the true predicate first for three consecutive emits while the aggregate has barely moved. AVG lift ranks it first after $9399$ tuples and never led incorrectly over $390$ pre-onset emits. $\Delta\Phi$ is the short horizon alerting signal; share--lift is the longer horizon explanatory one. Neither replaces a causal investigation.

\paragraph{Ad hoc mechanisms, measured.}
Table~\ref{tab:rq6_adhoc} instantiates Table~\ref{tab:adhoc} on NYC at $N=2\times10^5$. The retained state scan costs $\approx82$\,ms flat. The index costs $22\,\mu$s on the rarest predicate but rises to $15$\,ms as selectivity approaches $\Theta(N)$. Sampling answers in $2$--$4$\,ms independent of selectivity, with amortized ingest at $0.96\,\mu$s per tuple (below the index). At selectivity $1.5\times10^{-5}$, stratified sampling cuts error by $58\times$ versus a uniform sample that almost never matches. Rule of thumb: index what you can, scan when you must, and sample when the window is too large to retain.

\begin{table}[tbp]
\small
\setlength{\tabcolsep}{2.5pt}
\renewcommand{\arraystretch}{1.08}
\centering
\resizebox{\columnwidth}{!}{%
\begin{tabular}{|r|l|r|r|r|l|}
\hline
\textbf{sel.} & \textbf{mechanism} & \textbf{query ($\mu$s)} & \textbf{maint.} & \textbf{mem} & \textbf{rel.\ err} \\
 & & & \textbf{($\mu$s/t)} & \textbf{(MiB)} & \\ \hline
0.00002 & brute force scan & 80\,277.9 & 0.000 & 7.63 & 0 (exact) \\
 & index assisted & 22.4 & 1.470 & 3.05 & 0 (exact) \\
 & uniform sample & 4\,434.5 & 0.959 & 0.72 & 4.07e+01 \\
 & stratified sample & 2\,051.7 & 1.825 & 0.36 & 7.00e-01 \\
\hline
0.02515 & brute force scan & 82\,934.9 & 0.000 & 7.63 & 0 (exact) \\
 & index assisted & 1\,667.1 & 1.470 & 3.05 & 0 (exact) \\
 & uniform sample & 4\,354.4 & 0.959 & 0.72 & 1.45e+00 \\
 & stratified sample & 2\,087.2 & 1.825 & 0.36 & 1.18e+00 \\
\hline
0.62731 & brute force scan & 84\,649.4 & 0.000 & 7.63 & 0 (exact) \\
 & index assisted & 14\,894.7 & 1.470 & 3.05 & 0 (exact) \\
 & uniform sample & 3\,713.9 & 0.959 & 0.72 & 5.35e-01 \\
 & stratified sample & 2\,635.0 & 1.825 & 0.36 & 1.41e-01 \\
\hline
\end{tabular}%
}
\caption{RQ6, measured ad hoc mechanisms on \emph{nyc} at $N=2\times10^5$.}
\label{tab:rq6_adhoc}
\end{table}

\section{Related Work}

The companion paper~\cite{khani2026closedform} surveys the Shapley literature
that this work builds on. It covers stateless approximation, Shapley in
databases and data valuation, incremental XAI on model predictions,
explanation systems that are not axiomatic, and causality and provenance. Complementary prior work attributes static aggregates under partial causal graphs~\cite{khani2025causal,khani2026ccpi}; IDS instead maintains cooperative Shapley credit on sliding windows and does not claim causal diagnosis. We do not
repeat that survey. Instead, we discuss prior work specific to building a
\emph{runtime} on top of those results.

\paragraph{Sliding window aggregation engines.}
Classical pane frameworks~\cite{li2005pane,tangwongsan2015general} and modern
DSMS engines such as Apache Flink~\cite{carbone2015apache} evaluate
sliding window aggregates efficiently. The additivity of the IDS summaries makes
them compatible with those frameworks, but none of them expose predicate level
attribution, and none carry the objects IDS adds: an atom table, marginal
summaries maintained in parallel, optional retained expiry/index/sample state,
and a query surface in which attribution is a first class operator. Distributed
execution, fault tolerance, and production SQL integration remain outside this
paper's scope.

\paragraph{Sampling on sliding windows.}
This is the only mechanism that IDS needs but the companion paper does not use.
Classical reservoir sampling~\cite{vitter1985random} does not handle expiration.
IDS therefore builds on the chain-sample and priority-sample algorithms of
Babcock et al.~\cite{babcock2002sampling}. Their scheme realizes a size-$r$
sample as $r$ independent size-1 samplers and therefore samples \emph{with}
replacement. This design samples with replacement by default; Section~\ref{sec:rq4} measures the with- versus without-replacement distinction. Smooth histograms~\cite{braverman2007smooth}
offer an alternative space--precision tradeoff we do not currently exploit. The
concentration analysis uses Hoeffding~\cite{hoeffding1963probability},
Serfling~\cite{serfling1974probability}, and
Bardenet--Maillard~\cite{bardenet2015concentration}.

\paragraph{Stateless Shapley as a baseline.}
KernelSHAP~\cite{lundberg2017unified} and the permutation estimator of Castro et
al.~\cite{castro2009polynomial} treat each window as a fresh problem and form the
RQ6 baselines. Their per window cost is the quantity a maintained runtime is
supposed to eliminate, which is why they appear here as baselines rather than as
related methods.

\paragraph{Incremental Shapley under data change.}
The dynamic data valuation work of Zhang et al.~\cite{zhang2023dynamic} is
closest in spirit to IDS. It updates Shapley values under point insertions and
deletions instead of recomputing them from scratch. It targets model utility
games over a dataset and pays a per update cost that is more than constant. IDS instead
targets aggregate games over an expiring window and achieves $\mathcal{O}(1)$
evaluation from additive summaries. It treats expiry as a first class event
rather than as a deletion.

\section{Limitations and Future Work}

\textbf{Attribution is not causation.}
\texttt{SHAPLEY\_ATTRIBUTE} and \texttt{SHAPLEY\_DELTA} assign cooperative game credit. Ranking a planted predicate among candidates (Section~\ref{sec:rq6}) is an attribution and alerting result, not a physical root cause diagnosis.

\textbf{Standalone scope.}
IDS is evaluated as a single node library through deterministic offline replay. Distributed execution, fault tolerance, exactly once processing, and production DSMS integration are out of scope. Throughput figures use compiled kernels; the pure Python engine is slower and is the reference implementation.

\textbf{Aggregate scope.}
Shipped operators cover SUM, COUNT, AVG, and the two variance games, with a moment polynomial extension hook~\cite{khani2026closedform}. Quantiles, MIN/MAX, joins, and model inference queries remain out of scope.

\textbf{Retained state.}
Exact ad hoc scans, indexes, local expiry derivation, and exact promotion require $\mathcal{O}(N)$ retained active window state. Summaries-only deployments are possible when only registered predicates are queried and deltas are supplied externally.

\textbf{Out of order arrivals.}
The prototype assumes event-time ordered inserts and expirations within each emit tick. Late arrivals need reordering or retroactive correction of the affine form.

\textbf{Atom table fanout.}
Under many overlapping high-fanout predicates, $|\mathcal{S}_W|$ can approach $N$. Measured workloads kept $|\mathcal{S}_W|=12$ even at $K=64$ for mutually exclusive categoricals; adversarial schemas remain uncharacterized.

\FloatBarrier
\bibliographystyle{ACM-Reference-Format}
\bibliography{sample-base}
\end{document}